\pdfoutput=1
\documentclass[aps,prl,preprint]{revtex4}

\usepackage{graphicx}

\usepackage{ulem}
\usepackage{xcolor}

\begin{document}
\title{Implementation of
\\superconductor-ferromagnet-superconductor
$\pi$-shifters \\in superconducting digital and quantum circuits}


\author{A. K. Feofanov$^1$, V. A. Oboznov$^2$,
V. V. Bol'ginov$^2$, J. Lisenfeld$^1$, S. Poletto$^1$, V. V.
Ryazanov$^2$, A. N. Rossolenko$^2$, M. Khabipov$^3$, D. Balashov$^3$,
A. B. Zorin$^3$, P. N. Dmitriev$^4$, V. P. Koshelets$^4$,
and A. V. Ustinov$^1$}

\affiliation{
 $^1$ Physikalisches Institut and DFG Center for Functional Nanostructures (CFN),
 Karlsruhe Institute of Technology, Wolfgang-Gaede-Str.1, D-76131 Karlsruhe,
 Germany\\
 $^2$ Institute of Solid State Physics, Russian Academy of Science, Chernogolovka, 142432, Russia\\
 $^3$ Physikalisch-Technische Bundesanstalt, Bundesallee 100,
  38116 Braunschweig,   Germany\\
 $^4$ Kotel'nikov Institute of Radio Engineering and Electronics,
  Russian Academy of Science, Mokhovaya 11, Building 7,  Moscow, 125009, Russia
}

\begin{abstract}
The difference between the phases of superconducting order parameter
plays in superconducting circuits the role similar to that played by
the electrostatic potential difference required to drive a current
in conventional circuits. This fundamental property can be altered
by inserting in a superconducting circuit a particular type of weak
link, the so-called Josephson $\pi$-junction having inverted
current-phase relation and enabling a shift of the phase by $\pi$.
We demonstrate the operation of three superconducting circuits --
two of them are classical and one quantum -- which all utilize such
$\pi$-phase shifters realized using
superconductor-ferromagnet-superconductor sandwich technology. The
classical circuits are based on single-flux-quantum cells, which are
shown to be scalable and compatible with conventional niobium-based
superconducting electronics. The quantum circuit is a $\pi$-phase
biased qubit, for which we observe coherent Rabi oscillations and
compare the measured coherence time with that of conventional
superconducting phase qubits.
\end{abstract}

\maketitle

The fundamental property of superconducting weak links is a
$2\pi$-periodic current-phase relation. The supercurrent through a
conventional Josephson junction is usually described by the harmonic
relation $I_s=I_c \sin\varphi$, where $I_c$ is the critical current.
The so-called Josephson $\pi$-junction has the inverse current-phase
relation $I_s=I_c \sin(\varphi+\pi)=-I_c \sin\varphi$. The
$\pi$-junctions were theoretically proposed about three decades ago
\cite{Bul,Buz82}, whereas their remarkable properties have been
demonstrated in experiments notably later \cite{vanH7,Bas,Ryaz1}.
Practical implementations of $\pi$-junctions have been widely
discussed for a variety of different technologies. These include
approaches using superconductors with {\it d}-wave order parameter
symmetry \cite{vanH7,Testa-2004,Hilgenkamp-Nat-2003}, circuits with
nonequilibrium current injection \cite{Bas}, junctions with
ferromagnetic tunnel barriers \cite{Ryaz1}, and junctions with gated
carbon nanotubes \cite{Wernsdorfer-NatNano-2006}.

The ideas of using $\pi$-junctions in superconducting classical and
quantum circuits have been explored in several theoretical
proposals. In classical digital logic, a complementary Josephson
junction inverter \cite{Beasley} was suggested as a superconducting
analog of the complementary metal-oxide-semiconductor (CMOS) logic.
It relies on using Superconducting Quantum Interference Devices
(SQUIDs) of conventional ($0-$junctions) and $\pi$-types and
requires having similar parameters as $I_c$ and normal state
resistance for $0$- and $\pi$-junctions. These technologically
stringent requirements can be softened by using an alternative
"asymmetric" approach \cite{Ustinov} which employs $\pi$-junctions
as passive phase shifters (phase inverters) in basic cells of the
modified Single-Flux Quantum (SFQ) logic. Here the $\pi$-junction
critical current $I_c$ is chosen to be much larger than that of
conventional $0-$junctions employed in the very same SFQ cell, so
the phase difference across the $\pi$-junction is always close to
$\pi$ even in zero magnetic field. Since the total change of the
order parameter's phase over the closed path must become a multiple
of $2\pi$, the "missing" phase difference of $\pi$ or $-\pi$ is
induced on the rest part of the cell by a spontaneously generated
superconducting current.
A great advantage
of using such $\pi$ phase shifters is that the SFQ cell size can be
significantly reduced, opening the way to scaling superconducting
logic circuits down to sub-$\rm \mu$m dimensions \cite{Ustinov}.

The first proposal for using a loop with an integrated $\pi$-junction
as a superconducting quantum circuit \cite{ioffe99,blatter} featured
a superposition of two persistent current states in a loop at zero
magnetic field, in analogy to a spin-$1/2$ system. The
$\pi$-junctions required here must have very low dissipation (high
normal resistance), which so far has seemed unattainable for any of the
existing technologies for making $\pi$-junctions. The alternative
usage of $\pi$-junctions as passive phase shifters offers an
advantage for the operation of superconducting flux qubits at the
degeneracy point requiring zero or a very small external magnetic
field. Potentially, this allows noise and
electromagnetic interference induced by magnetic field sources
to be minimized.
There remains an open question: Do $\pi$-junctions themselves
introduce any intrinsic decoherence when they are inserted into a
superconduction quantum circuit?

Here we describe first practical implementation and operation of
$\pi$-shifters in digital and quantum circuits. $\pi$-junctions that
we use are based on superconductor-ferromagnet-superconductor (SFS)
Josephson  multilayer technology. The origin of the $\pi$-state in an
SFS junction is an oscillating and sign-reversing superconducting
order parameter in the ferromagnet close to the SF interface
\cite{Buz82,Buz_rev}. Due to these oscillations, different signs of
the order parameter can occur at the two banks of the SFS sandwich
when the F-layer thickness is of the order of half an oscillation
period, which corresponds to a sign change of the supercurrent and a
negative Josephson coupling energy. This behavior was first observed
experimentally on Nb-CuNi-Nb sandwiches in Ref.~\cite{Ryaz1}.
Further experiments reported the spontaneous flux
\cite{Frolov-NatPhys-2008} and half-periodical shifts of the
superconducting interferometer $I_c(H)$ dependence \cite{Ryaz2} as
well as a sign change of the junction current-phase relation
\cite{Frolov}. Recently, the critical current density of the
Nb-Cu$_{0.47}$Ni$_{0.53}$-Nb $\pi$-junctions was pushed above 1000
A/cm$^2$ \cite{Oboznov}. These junctions are compatible with
conventional niobium thin film technology and thus can be easily
integrated in the conventional fabrication process of
superconducting digital circuits.

To verify the operation of $\pi$-junction phase shifters in an
analog regime, we fabricated two geometrically identical
superconducting loops on a single Si substrate (see Fig.~1(c))
schematically shown in Figs.~1(a) and (b). Circuit (b) is a
two-junction interferometer conventionally called a DC-SQUID.
Configuration of circuit (a) is nominally identical to (b), except
that an SFS $\pi$-junction has been additionally inserted in the
left branch of the loop seen in the lower left corner of the circuit
image in Fig.~1(c). The on-chip distance between the centers of the
two loops is 140 $\rm \mu$m, so both interferometers are exposed to
the same magnetic field during the experiment. Figure~1(d) shows the
schematic cross-section of a thin-film Nb-CuNi-Nb sandwich forming
the $\pi$-junction, which fabrication procedure is described in
supplementary material. The junction normal resistance $R_n$ is
about 150 $\rm \mu \Omega$ at an F-layer thickness of 15 nm. The
critical currents of such $\pi$-junctions are about 200 $\mu$A and
hence the junctions do not switch to the resistive state when
embedded in loops with conventional tunnel junctions having critical
currents of about 10 $\mu$A. This large difference between two
critical currents means that during the dynamic switchings in the
rest of the circuit, $\pi$-junctions do not introduce any noticeable
phase shifts deviating from $\pi$.

\begin{figure}
    \centering
    \includegraphics[width=85mm, clip] {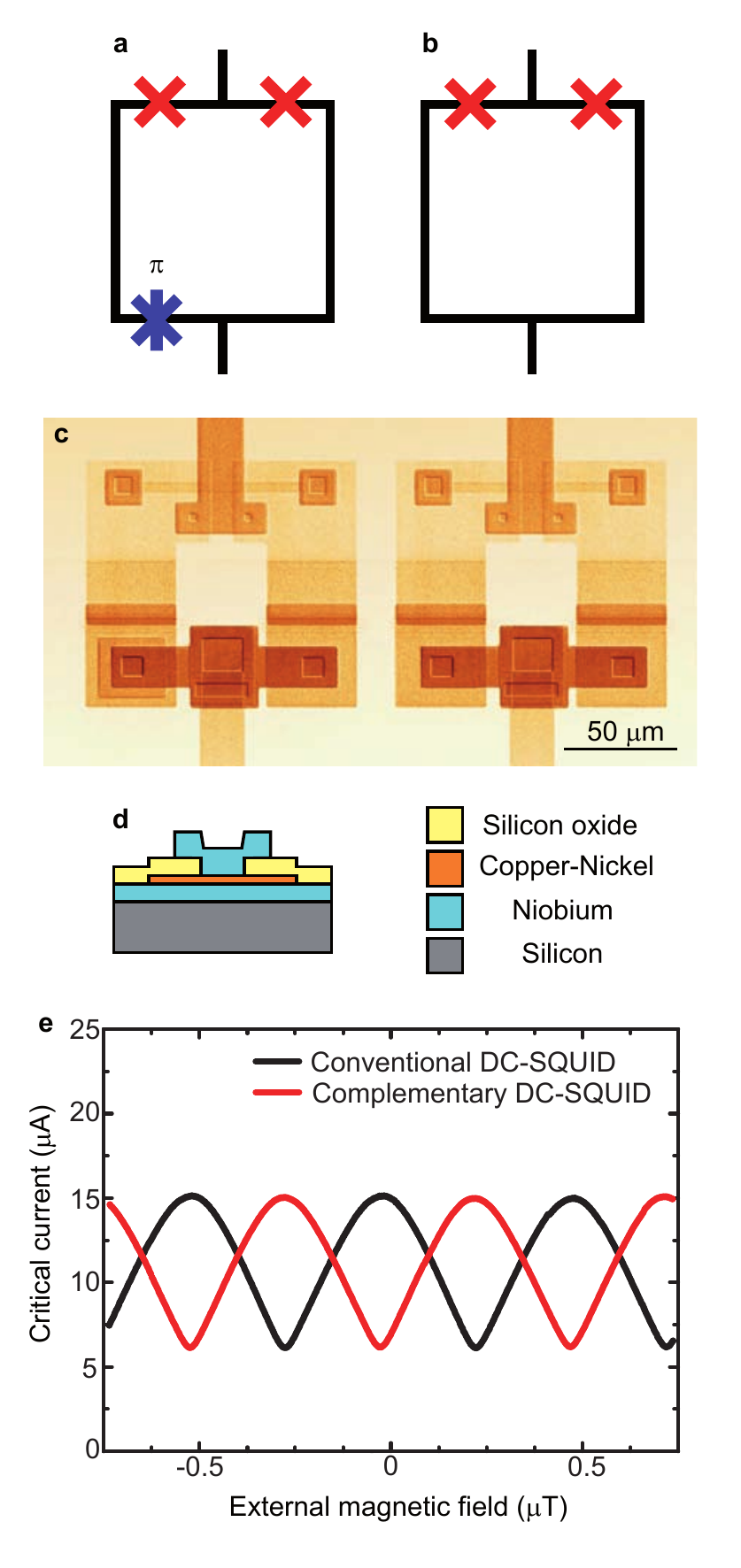}
    \caption{Complementary DC-SQUIDs. {\bf a}, Schematic of a
complementary DC-SQUID employing two conventional Josephson
junctions (red crosses) and a $\pi$-junction (blue star). {\bf b},
Schematic of a conventional DC-SQUID used as a reference device.
{\bf c}, A SEM micrograph of the fabricated DC-SQUIDs. The
ferromagnetic layer is visible in the lower left corner of the left
device. {\bf d}, Schematic cross-section through an SFS
$\pi$-junction. {\bf e}, Dependencies of the critical currents of
the devices shown in {\bf c} vs. the applied magnetic flux. The red
curve related to the $\pi$-SQUID is shifted by half a period.
} \label{fig1}
\end{figure}

The dependencies of the critical currents $I_c(H)$ of the two
devices shown in Figs.~1(a) and (b) are presented in Fig.~1(e).
Whereas both curves have the same shape, they are shifted by a
half-period. A small offset of the symmetry axes for both curves
from the zero-field value is due to a residual magnetic field in the
cryostat. The minimum of the red $I_c(H)$ curve at zero field is
%
due to inclusion of the $\pi$-junction in the superconducting loop.
In the conventional SQUID the same frustrated state exists at an
external magnetic field corresponding to half-integer numbers of
magnetic flux quanta per cell. Thus, embedding an SFS $\pi$ phase
shifter into a superconducting loop indeed leads to self-biasing of
the loop by a spontaneously induced supercurrent.

In the second experiment, we demonstrate the functionality of the
$\pi$ phase shifter included in a superconducting logic circuit. The
SFQ logic circuits enable processing of information in the form of
single flux quanta which can be stored in elementary superconducting
cells including inductors and Josephson junctions. Dynamically, this
information is represented by SFQ voltage pulses
\cite{LikharevSemenov} having quantized a area $\int V(t)dt =
\Phi_0$ and corresponding to the transfer of one flux quantum across
a Josephson junction. The first SFQ circuits with active $\pi$
elements were made of high-$T_c$ superconductor
(YBa$_{2}$Cu$_{3}$O$_{7-\delta}$) employing  $d$-wave pairing
symmetry combined with conventional low-$T_c$ superconductor (Nb)
\cite{Ortlepp}. Operation of the circuits with the phase shifting
element based on frozen flux quanta \cite{Majer} has been tested
earlier in Ref.~\cite{Balashov}. Here we present the first
demonstration of the functioning of an SFS $\pi$-phase shifter
integrated in a conventional Nb SFQ circuit.

\begin{figure}
    \centering
    \includegraphics[width=155mm, clip] {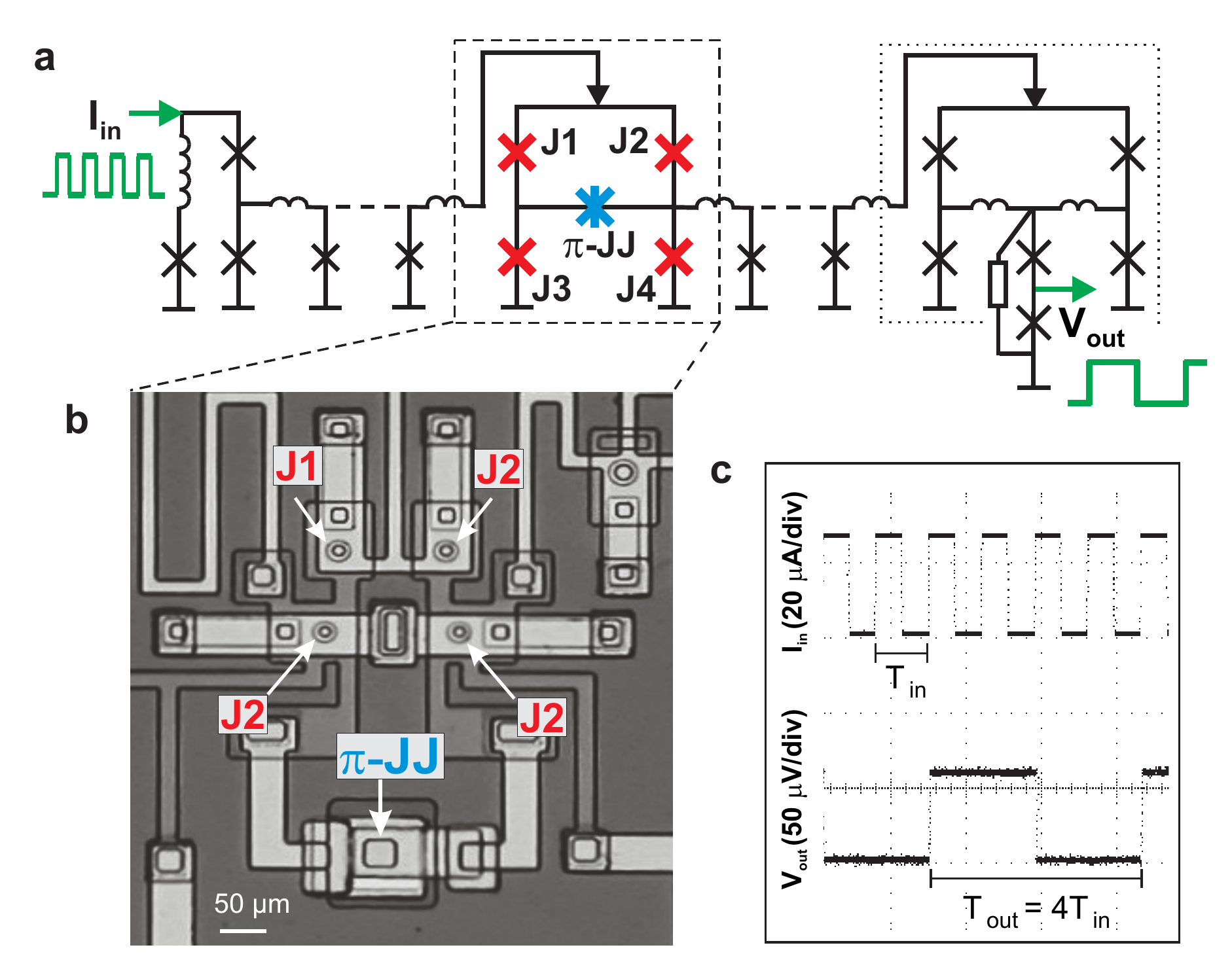}
    \caption{{\bf a}, Schematic of the frequency
    binary divider. This two-stage circuit includes the
    Toggle Flip-Flop (TFF) with SFS $\pi$-junction (inside the dash-line box) and
      conventional TFF (inside dotted-line box), with Josephson transmission lines
      enabling delivery of SFQ pulses to the inputs of TFFs.
    {\bf b}, Micrograph of these TFF with $\pi$-junction. {\bf c},
Oscilloscope output verifying correct operation of this SFQ circuit, i.e.
division of the frequency of input pulses by four.} \label{fig2}
\end{figure}

Figure 2 shows the layout and operation of our test SFQ circuit,
represented by a two-stage frequency divider. The positive edges of
rectangular trigger pulses ($I_{\rm{in}}$) are delivered to the
first dividing stage of the circuit via a Josephson transmission
line (left part of diagram in Fig.~2(a)) in the form of SFQ pulses.
These pulses cause switching of this first stage of the circuit, the
Toggle Flip-Flop (TFF) with an integrated SFS $\pi$-phase shifter
which replaces a large inductance required in the conventional
counterpart circuit for the realization of desired bistable
behavior. The output signal of the first TFF in the form of SFQ
pulses is sent to the second dividing stage which is realized as a
conventional TFF circuit. The circuit output $V_{\rm{out}}$ time
pattern shows a division of the frequency of the input pulses by
four (with rather large margins for circuit parameters), which
verifies its correct functioning.

Another attractive application of SFS $\pi$-junctions is their use
as phase shifters in coherent quantum circuits realizing
superconducting quantum bits. The answer to the question of whether
or not $\pi$-junctions can become useful in superconducting circuits
designed for quantum computing applications depends on their impact
on the coherence properties of the qubits. Potential sources of
decoherence introduced by $\pi$-junctions can for instance be
spin-flips in the ferromagnetic barrier, either occurring randomly
or being driven by high-frequency currents and fields, as well as
the dynamic response of the magnetic domain structure. We address
these important coherence issues in a third experiment reported in
this paper, in which we use an SFS $\pi$-junction to self-bias a
superconducting phase qubit. We have chosen here a phase qubit
\cite{Simmonds} rather than a flux qubit \cite{Mooij-03} due to the
simpler fabrication procedure for the former. The results reported
below would nevertheless remain fully applicable to flux qubits.

\begin{figure}
    \centering
    \includegraphics[width=85mm, clip] {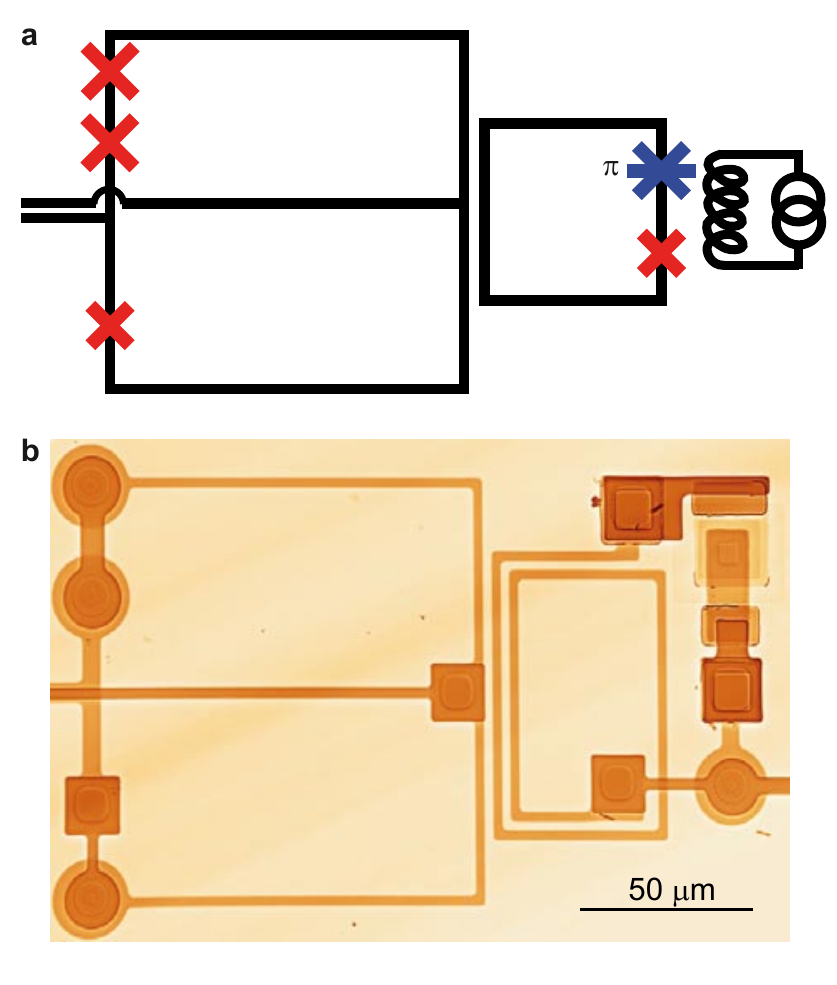}
    \caption{Self-biased phase qubit. {\bf a}, Schematic of
 a phase qubit circuit used to test the decoherence
properties of the $\pi$-junction. The qubit is realized by the
central loop with embedded conventional and $\pi$ - Josephson
junctions. The larger loop to its right is a DC-SQUID for qubit
readout. A current-biased coil coupled weakly to the qubit is used
for flux-biasing the qubit. {\bf b}, SEM picture of the realized
phase qubit employing a $\pi$ - junction in the qubit loop.}
\label{fig3}
\end{figure}

A phase qubit \cite{Simmonds} consists of a single Josephson
junction embedded in a superconducting loop. It is magnetically
biased close to an integer number of flux quanta in the loop. At
such a bias, the potential energy of the qubit exhibits an
asymmetric double-well potential, whereas two quantized energy
eigenvalues of the phase localized inside only the shallow well are
used as the logical qubit states $|0\rangle$ and $|1\rangle$. The
qubit is controlled by inducing a small-amplitude microwave current
in the loop whose frequency is tuned in resonance to the $|0\rangle$
to $|1\rangle$ transition, giving rise to Rabi oscillation of the
state population. Reading out the qubit is accomplished by applying
a short dc flux pulse to the qubit loop, during which only the
excited qubit state may tunnel to the neighboring potential well.
Since this tunneling event entails a flux quantum entering the qubit
loop, reading out the qubit is concluded by a measurement of the
flux threading the qubit loop by means of an inductively coupled
DC-SQUID. Figure 3(a) shows a circuit schematic and 3(b) a micrograph
of the tested sample. Here, a $\pi$-junction is connected in series
to the phase qubit's tunnel junction. This circuit was fabricated in a
standard Nb/Al-AlO$_x$/Nb trilayer process, whereas the
$\pi$-junction was integrated subsequently by performing the
additional lithographic steps described above. Coherent qubit
operation is demonstrated by the data reported in Fig.~4(a), showing
Rabi oscillation of the excited qubit state population probability
in dependence on the duration of a resonant microwave pulse. Each
data set was taken using the indicated microwave power as delivered
by the generator, giving rise to a change in the coherent
oscillation frequency as expected for Rabi oscillation. The
oscillations exhibit a decay time of about 4 ns, which is a typical
value reachable in samples fabricated using similar fabrication
processes \cite{Lisenfeld}. To find out whether $\pi$-junction does
introduce additional decoherence, a conventional phase qubit without
a $\pi$-junction was fabricated on the same wafer. This qubit,
however, required a constant phase bias of about $\pi$ set by
applying an external magnetic field. As shown in Fig.~4(b),
this reference qubit shows a nearly identical decay time for Rabi
oscillations, allowing us to conclude that at least on the
observable time scale no extra decoherence is introduced by the SFS
$\pi$ phase shifter employed in this circuit.

\begin{figure}
    \centering
    \includegraphics[width=85mm, clip] {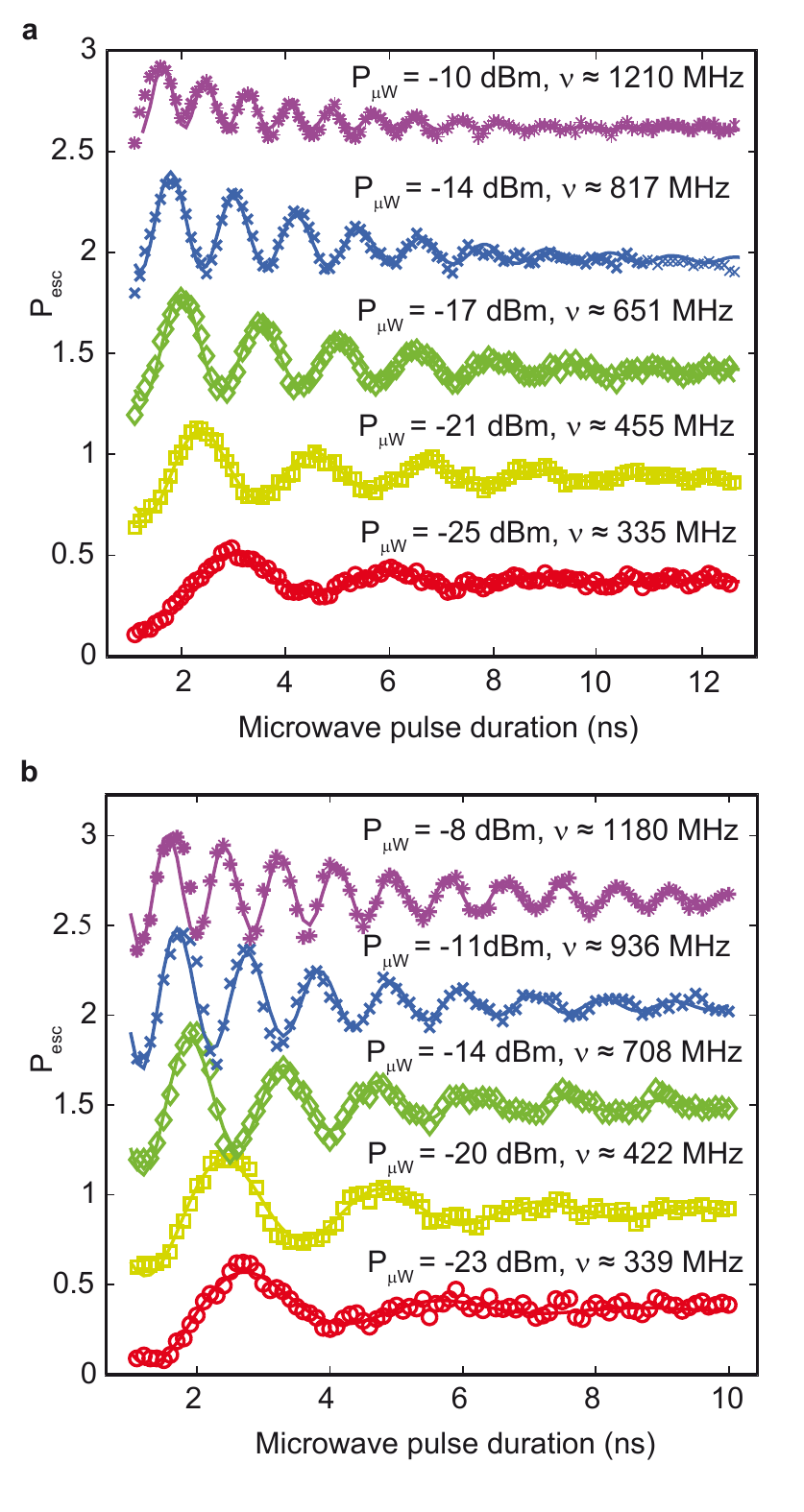}
    \caption{Rabi oscillation of the occupation probability of
    the excited qubit state resulted from resonant microwave driving.
    {\bf a}, Observed in the phase qubit
with embedded $\pi$-junction, and {\bf b}, A conventional phase
qubit made on the same wafer as a reference.} \label{fig4}
\end{figure}

We compared the measured decoherence time with the theoretical
predictions \cite{Kato-Golubov-2007}. We assume here an
\textit{overdamped} SFS $\pi$-junction having a normal resistance of
$R_{N,\pi}\approx$ 500~$\rm \mu\Omega$ and a critical current
$I_{C,\pi}\approx$ 50~$\mu$A. In our case, the qubit level splitting
$\Delta \gg 2 eI_{C,\pi}R_{N,\pi}$, where $\Delta\approx h\cdot$
13.5 GHz, $h$ is the Plank's constant and $e$ is the elementary
charge. Here, the energy $2 eI_{C,\pi}R_{N,\pi}\approx h\cdot
12\,$MHz is associated with characteristic Josephson frequency of
our SFS $\pi$-junction. Simplifying the expression for the
relaxation time~\cite{Kato-Golubov-2007} in this limit (see
supplementary material), we can theoretically estimate the
relaxation time $\tau_{\rm relax}$ as

\begin{equation}
\tau_{\rm relax}\approx \frac{\Delta}{2 I^2_C R_{N,\pi}}\approx
2\,{\rm ns}.
\end{equation}
Here, $I_C\approx$ 2 $\mu$A is the critical current of the small SIS
qubit junction. The estimated value of the energy relaxation time is
in good agreement with the measured decoherence time. We note,
however, that the relaxation time (1) can be significantly enhanced
by using SFS junctions with a smaller resistance $R_{N,\pi}$.

In contrast to $\pi$-junctions based on high-$T_c$ superconductor
junctions with $d$-wave pairing symmetry, SFS elements can have
a sufficiently large critical current, so the desired $\pi$ phase
shift remains reliably fixed during circuit operation. In
distinction from phase-shifting loops with frozen magnetic flux
\cite{Majer}, the SFS circuits are much more compact and do not
require trapping a well-defined integer number flux quanta in their
superconducting loops.

As an outlook, a significant reduction in the size of the
demonstrated SFS $\pi$-phase shifters for digital circuits is
readily possible. The visualization of the magnetic structure of our
F layer material shows domain sizes smaller than 100 nm. Therefore,
we believe that a reduction of the junction planar dimensions down
to 300-500 nm is feasible. Furthermore, combining the high-$j_C$
$\pi$-junction technology with in-situ grown tunnel barriers
\cite{Weides1,Weides2} may open the way towards active inverter
elements which are in great demand for superconducting electronics.

In summary, we demonstrated here the successful operation of three
generic superconducting circuits with embedded $\pi$-junction phase
shifters. For the studied $\pi$-biased phase qubit, we observed Rabi
oscillations and compared their coherence time with that of
conventional phase qubits fabricated by the same technology. We find
no degradation of the coherence time induced by the presence of the
$\pi$-junction. The demonstrated SFS $\pi$-junction phase shifter
circuits are readily scalable and compatible with conventional
niobium-based superconducting circuit technology.


\section{Acknowledgements}
This work was supported by the EU projects EuroSQIP and MIDAS. We
acknowledge support by the Deutsche Forschungsgemeinschaft (DFG),
the joint grant of DFG and Russian Foundation of Basic Research, the
Russian Federal Agency of Science and Innovations, and the State of
Baden-W\"urttemberg through the DFG Center for Functional
Nanostructures (CFN).

\section{Competing Interests}
The authors declare that they have no competing financial interests.

\section{Correspondence}
Correspondence and requests for materials should be addressed to
A.V. Ustinov~(email: ustinov@physik.uni-karlsruhe.de).

\section{Supplementary materials}

\subsection{Sample fabrication}

For fabrication of SFS $\pi$-junctions, the bottom Nb-electrode with thickness of 110 nm was fabricated by dc-magnetron sputtering followed by a lift-off process. The deposition of the copper-nickel layer
(F layer) was carried out by rf sputtering after ion cleaning of the
niobium surface. Afterwards, the insulating layer having a
10$\times$10 $\rm \mu m^2$ window which determines the junction area
was prepared by the lift-off process. We used a 150 nm thick SiO film as
insulator, which was thermally evaporated. The fabrication procedure
was completed by Ar plasma cleaning and dc-magnetron sputtering of
the upper niobium electrode of 240 nm thickness.

Details on the fabrication technique for tunnel Nb/Al/AlO$_x$/Nb
junctions employed in this circuit are presented in
Ref.~\cite{Koshelets-tech}. In brief, a three-layer Nb/Al/AlO$_x$/Nb
structure is deposited by magnetron sputtering. The layers have
thicknesses of 180, 7, and 80 nm, respectively. Aluminum is oxidized
in pure oxygen to form a tunnel barrier having a critical current
density of about 200 A/cm$^2$. The junction area, here 10 $\rm
\mu$m$^2$, is defined by reactive ion etching and subsequent SiO$_2$
deposition. Resistive shunts in parallel to the tunnel junctions are
formed by a molybdenum layer with a specific resistance of 2 $\rm
\Omega$ per square.

\subsection{Estimates of decoherence in $\pi$-junctions}

In the paper by Kato, Golubov and Nakamura \cite{Kato-Golubov-2007},
the following expressions for the effective noise spectrum
$J_{\rm eff}$ and  relaxation time $\tau_{\rm relax}$ were obtained:
\begin{equation}
J_{\rm eff}(\omega)=\frac{8E^2_{J}E_{C,\pi}}{\hbar^3}\left[
\frac{\gamma \omega}{\gamma^2\omega^2 +
(\omega^2-\omega^2_0)^2}\right],
\end{equation}
where $\gamma = \frac{1}{R_\pi C_{\pi}}$, $\omega_0=\frac
{\sqrt{8E_{J,\pi}E_{C,\pi}}}{\hbar}$, $E_{C,\pi}=\frac{e^2}{2C_\pi}$
and
\begin{equation}
\tau^{-1}_{\rm relax}=2J_{\rm
eff}(\Delta/\hbar)\coth\left({\frac{\Delta}{2k_BT}}\right),
\end{equation}
where $\Delta$ is the qubit level splitting. These equations were
derived using an RSJ model under an assumption $E_{J,\pi}\gg E_{J}\gg
E_C$, which means the $\pi$-junction remains in the superconducting
state and we are working in the regime of a well defined phase.

Assuming that $\omega \ll \omega_0$, we can write
\begin{eqnarray}
J_{\rm eff}(\omega)\approx \frac{8E^2_{J}E_{C,\pi}}{\hbar^3}\left[
\frac{\gamma
\omega}{\gamma^2\omega^2 + \omega^4_0}\right]=\\
=\frac{8E^2_{J}E_{C,\pi}}{\hbar^3}\left[ \frac{\gamma
\omega}{\gamma^2\omega^2 +
64\frac{E^2_{J,\pi}E^2_{C,\pi}}{\hbar^4}}\right]=\\
=\frac{4E^2_{J}\frac{e^2}{R_{N,\pi}
C^2_\pi}\omega}{\frac{\hbar^3}{R^2_{N,\pi} C^2_{\pi}}\left[ \omega^2
+ 16\frac{E^2_{J,\pi}}{\hbar^4}e^4 R^2_{N,\pi}\right]}.
\end{eqnarray}
Now we can exclude $C_\pi\,$,
\begin{equation}
J_{\rm eff}(\omega)\approx\frac{4E^2_{J}e^2 R_{N,\pi}
\omega}{\hbar^3\left[ \omega^2 + 16\frac{E^2_{J,\pi}}{\hbar^4}e^4
R^2_{N,\pi}\right]}.
\end{equation}
Since $E_J=\frac{I_C\Phi_0}{2\pi}=\frac{I_C\hbar}{2e}$,
it allows us to further simplify the latter expression:
\begin{eqnarray}
J_{\rm eff}(\omega)\approx\frac{4\frac{I^2_C\hbar^2}{4e^2}e^2
R_{N,\pi} \omega}{\hbar^3\left[ \omega^2 +
16\frac{I^2_{C,\pi}\hbar^2}{4e^2}\frac{e^4}{\hbar^4}R^2_{N,\pi}\right]}=\\
=\frac{I^2_C R_{N,\pi} \hbar\omega}{\hbar^2\omega^2 +
4e^2I^2_{C,\pi}R^2_{N,\pi}}.
\end{eqnarray}
The relaxation time can be calculated as
\begin{eqnarray}
\tau_{\rm relax}= \left[2J_{\rm eff}(\Delta/\hbar)\coth\left({\frac{\Delta}{2k_BT}}\right)\right]^{-1}\approx\\
\approx\left[2\left(\frac{I^2_C R_{N,\pi}
\hbar\frac{\Delta}{\hbar}}{\hbar^2\left(\frac{\Delta}{\hbar}\right)^2
+
4e^2I^2_{C,\pi}R^2_{N,\pi}}\right)\coth\left({\frac{\Delta}{2k_BT}}\right)\right]^{-1}=\\
=\left(\frac{\Delta^2+ 4e^2I^2_{C,\pi}R^2_{N,\pi}}{2I^2_C R_{N,\pi}
\Delta}\right)\tanh\left({\frac{\Delta}{2k_BT}}\right).
\end{eqnarray}
If the qubit level splitting $\Delta \gg 2 eI_{C,\pi}R_{N,\pi}$ we
can neglect the second term in the numerator:
\begin{equation}
\tau_{\rm relax}\approx\left(\frac{\Delta}{2I^2_C
R_{N,\pi}}\right)\tanh\left({\frac{\Delta}{2k_BT}}\right)\approx
2\,{\rm ns.}
\end{equation}
In our case $\Delta\approx h\cdot 13.5\,$GHz, $I_C\approx 2\,\mu$A,
$R_{N,\pi}\approx 500\,{\rm \mu\Omega}$. The temperature $T$ was
about 50 mK, thus
$\tanh\left({\frac{\Delta}{2k_BT}}\right)\approx$~1.

\end{document}